\documentclass[a4paper,twocolumn]{esapub2005}
\usepackage{times}
\usepackage{natbib}

\usepackage{graphicx}
\usepackage{epsfig}

\title{Eclipsing Binary System WN3(h)+O5\,V BAT99-129: Analysis of the MACHO
Light Curve and the Parameters of the Components.}

\author{I.I.Antokhin, A.M.Cherepashchuk \\ Sternberg Astronomical Institute}

\begin{document}

\maketitle

\abstract{

BAT99-129 is a massive eclipsing binary system in the Large Magellanic Cloud
(LMC) which consists of WN3(h) and O5\,V components. A broad-band MACHO
light curve of the system is studied in the present paper. A dense and
extended atmosphere of the Wolf-Rayet (WR) star does not allow one to
analyze the light curve in terms of standard parametric models of
Wilson-Devinney type. Distributions of brightness and absorption across the
WR star disk are restored using direct solution of integral equations
describing eclipses in the system.

As a result, reliable estimates of the orbital inclination and component
parameters have been obtained. The orbital inclination is $78^\circ$, the
orbital separation is $28.5\,R_\odot$, the radius of the O component is
$R_{\rm O}=7.1\,R_\odot$. The size of the WR component core which is opaque
in optical continuum is $R_{\rm WR}=3.4\,R_\odot$. The brightness
temperature in the center of the WR disk is $T_{\rm br}=45\,000$\,K.
Probable errors of the parameters are discussed. Evolutionary status of the
system is discussed.

}

\section{Introduction}

Wolf-Rayet stars (WR) are considered to be helium remnants of evolved
massive stars with initial masses of some dozens solar masses. A WR stage in
the life of a massive star does not last very long. Nevertheless, these
objects play a key role in understanding evolution and internal structure of
massive stars because of combination of unique physical characteristics
which make WR stars very valuable objects for comparison between theory and
observations. One of the prominent features of WR stars is their strong
stellar winds (mass loss rate may approach $\sim10^{-4}{\rm M_\odot}$/year),
the origin of which is not fully clear yet. Dense winds make it very
difficult to determine physical parameters of the central WR star cores.
For example, two important parameters are the effective temperature of the
WR star and the radius of its opaque core. This radius can be defined as a
radius where the radial Rosseland optical depth is equal to 2/3
\cite{hamann04}. When the luminosity is known, the effective temperature of
a star can be obtained using Stefan-Boltsman law.

In principle, main characteristics of single WR stars (temperature, radius,
and mass loss rate $\dot M$) can be determined from comparison between their
theoretical and observational spectra. For dense winds, however, which are
characteristic of early WN subclasses there is a correlation \cite{hamann04}
between model parameters. The correlation allows only the combination of
parameters $L/{\dot M}^{4/3}$ to be determined where
$L=4\pi{R_*^2}\sigma{T_{\rm eff}^4}$ is the luminosity of the star,
$R_*$ is its radius, $T_{\rm eff}$ is its effective temperature,
$\sigma$ is the Stefan-Boltsman constant.

Eclipsing binary systems containing a WR component provide a possibility
to directly determine the radius of the opaque core and the brightness
temperature of the WR by means of light curve analysis.
A ``normal'' component in the system serves as a probe object. A strong
semi-transparent wind from the WR makes parametric modeling in binary
systems (Russell-Merrill type \cite{russell52}, Wilson-Devinney type
\cite{wilson79}, etc.) rather problematic. In this paper we use another
approach, namely, direct solution of integral equations describing light
losses during eclipses.

The second section of the paper contains the basic information on BAT99-129
and observational data which are used in our work. The third section
describes the method of light curve solution. The forth section provides the
results of the light curve solution for BAT99-129. The fifth section
presents a discussion of the results obtained and summarizes the
conclusions of the work.

\section{Information on the system. Observational data }

Eclipsing binary system BAT99-129 in the LMC is one of a few known
extragalactic WR eclipsing binary systems. Spectral observations of this
star were recently analyzed in detail by \cite{foellmi06}. Its spectral
class was found to be  WN3(h)\,+\,O5\,V and the orbital period
$P=2^d.7689\pm{0^d.0002}$. The orbit of the binary system seems to be
circular (zero eccentricity was accepted in the paper cited). The orbital
separation is $a\sin{i}=27.9\,R_\odot$. The ratio of the optical
luminosities of the components has been determined using two
spectrophotometric methods. The mean value of the ratio is $L^{obs}_{\rm
WR}/L^{obs}_{\rm O}\,=\,0.34\pm{0.2}$. The radii and effective temperatures
of the O and WR components \cite{foellmi06} were estimated approximately
from their evolutionary status and corresponding model parameters. The
effective temperature of the O component was accepted to be
$T_{eff}=43\,000$\,K while its radius is comprised between 8.8 and
10.5\,$R_\odot$. In reality, the O star may belong to O6 subclass as
evidenced by some spectral features, thus having a smaller radius. The
temperature of the WR component was estimated to be $71\,000$\,K and the
radius, which corresponds to the Rosseland optical depth of 20,
$R_*=4.7\,R_\odot$. These values of radii as well as the supposed component
masses allowed the authors to conclude that the inclination angle of the
orbit may be around 60$^\circ$ or more.

BAT99-129 is an object in the sight of MACHO (Massive Compact Halo
Objects)\footnote{The home page of the experiment is located at
http://wwwmacho.anu.edu.au.}. In the course of the experiment a long-term
photometric monitoring of a few million objects in the LMC and Milky Way is
being carried out. The data are acqured in two color bands prodived by a
dichroic filter. The ``blue'' band is sensitive to emission between
4500\,--\,5900\,\AA while the ``red'' one covers the range 5900\,--\,7800\,\AA.

\begin{figure}
\epsfig{file=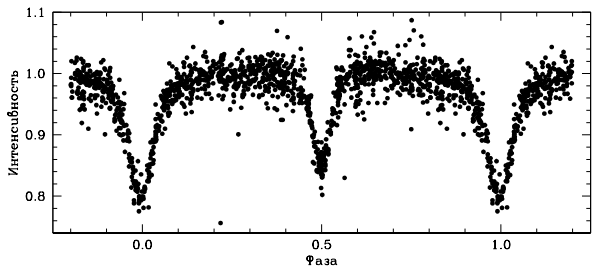,width=\linewidth}

\caption{The light curve of BAT99-129 obtained in the MACHO experiment,
individual measurements. The curve is reduced to relative intensities
normalized to the maximum. The latter is determined as mean intensity
between the phases 0.3 and 0.35.}

\label{fig-macho_lc}
\end{figure}

The data from more than seven years of observations for BAT99-129 system are
available in both bands since 1992. A total number of individual
measurements reaches 877 in the blue band and 461 in the red one. After
having been analyzed, the corresponding light curves do not show any
significant differences in their shape. Hence, to analyze the light curve,
the data in both bands have been combined and a joint light curve has been
analyzed instead. The individual measurements for BAT99-129 convoluted with
the above mentioned period and the initial epoch
$E_0$\,[HJD]\,=\,2448843.8935 taken from \cite{foellmi06} are shown in
Fig.\,\ref{fig-macho_lc}.

The light curve in Fig.\,\ref{fig-macho_lc} is symmetrical with respect to
the phase 0.5. For this reason, to carry out further analysis, the right
part of the light curve (the phases 0.5\,--\,1.0) has been reflected
symmetrically relative to the phase 0.5. After dropped-out points were
removed and individual measurements were averaged, a mean light curve has
been obtained in the phase interval 0.0\,--\,0.5. Intervals for the
averaging were selected in such a manner that the mean light curve could
describe light variability in the most optimal way. A mean-square error of a
normal point in the mean light curve lies between $0.002$ and $0.005$.
Normal points of the mean light curve together with their errors are given
in Table\,\ref{table_obsdata}. The table includes both non-rectified and
rectified light curves (see below).

\begin{table}[t]
\caption{Mean Light Curve of BAT99-129. $1-l$ is the light loss at a given orbital phase.}
\label{table_obsdata}
\centering
\begin{tabular}{|r|r|r|r|r|}
\hline
\multicolumn{1}{|c|}{$\theta$} & \multicolumn{1}{|c|}{1-l}  & \multicolumn{1}{|c|}{$\sigma$} & \multicolumn{1}{|c|}{1-l} & \multicolumn{1}{|c|}{$\sigma$} \\
\hline
\multicolumn{5}{|c|}{Primary minimum} \\
\hline
 & \multicolumn{2}{|c|}{nonrectified} & \multicolumn{2}{|c|}{rectified} \\
 2$^\circ$.1528 &  0.1992 & 0.0024  &   0.1807 & 0.0025 \\
 5.9982 &  0.1779 & 0.0035  &   0.1590 & 0.0036 \\
 9.4967 &  0.1497 & 0.0063  &   0.1304 & 0.0064 \\
12.9269 &  0.1169 & 0.0058  &   0.0972 & 0.0059 \\
16.0841 &  0.0874 & 0.0035  &   0.0674 & 0.0036 \\
19.8393 &  0.0611 & 0.0049  &   0.0411 & 0.0050 \\
24.5390 &  0.0529 & 0.0024  &   0.0336 & 0.0024 \\
32.5004 &  0.0262 & 0.0021  &   0.0082 & 0.0021 \\
44.7968 &  0.0160 & 0.0017  &   0.0012 & 0.0017 \\
62.3288 &  0.0082 & 0.0016  &  -0.0012 & 0.0016 \\
81.3507 &  0.0018 & 0.0015  &  -0.0024 & 0.0015 \\
\hline
\multicolumn{5}{|c|}{Secondary minimum} \\
\hline
 0$^\circ$.5667 &  0.1495 & 0.0049  &   0.1491 & 0.0049 \\
 2.7342 &  0.1405 & 0.0031  &   0.1401 & 0.0031 \\
 5.0691 &  0.1178 & 0.0029  &   0.1174 & 0.0029 \\
 9.0434 &  0.0919 & 0.0029  &   0.0916 & 0.0029 \\
12.5372 &  0.0575 & 0.0029  &   0.0573 & 0.0029 \\
16.2047 &  0.0244 & 0.0038  &   0.0243 & 0.0038 \\
23.1942 &  0.0091 & 0.0027  &   0.0092 & 0.0027 \\
33.1889 &  0.0055 & 0.0031  &   0.0060 & 0.0031 \\
44.8749 & -0.0004 & 0.0014  &   0.0005 & 0.0014 \\
62.8766 & -0.0038 & 0.0015  &  -0.0029 & 0.0015 \\
81.3796 &  0.0056 & 0.0015  &   0.0048 & 0.0015 \\

\hline

\end{tabular}
\end{table}

\section{Method for light curve solution}

To solve the light curve of BAT99-129 we used the method that was described in detail in
\cite{ant97} and \cite{ant01}. Hence only its basic features are briefly
described here and those who are interested in details are referred to the
papers cited.

Let us suppose that both components of an eclipsing binary system are
spherical and the brightness/opacity distributions over their disks are
axially symmetric. Then the light curve of the system containing a component
with an extended atmosphere could be described by the following system of
integral and algebraic equations:

{\small
\begin{equation}\label{eq1}
\begin{array}{rlr}
1-l_1(\theta)= & \int\limits_0^{R_a}K_1(\xi,\Delta,R_{\rm O})I_0I_a(\xi){\rm d}\xi & (a)\\
1-l_2(\theta)= & \int\limits_0^{R_c}K_2(\xi,\Delta,R_{\rm O})I_c(\xi){\rm d}\xi & (b)\\
\multicolumn{2}{c}{ L_{\rm WR}+L_{\rm O}=I_0\pi{R_{\rm O}^2}(1-\displaystyle\frac{x}{3})+\int\limits_0^{R_c}I_c(\xi)2\pi\xi{\rm d}\xi=1} &
\end{array}
\end{equation}
}

\noindent Here $1-l_{1,2}(\theta)$ are observational light losses in the
primary and secondary minima respectively, $\theta$ is the relative
positional angle between the components connected with the distance $\Delta$
between the centers of the component disks by the formula
$\Delta^2=\cos^2i+\sin^2i\,\sin^2\theta$, (the orbital separation 
and the total luminosity of the system are suggested to be equal to unity),
$R_{\rm O}$ is the radius of the O star, $R_c$ and $R_a$ are the upper
limits for the total radii of WR emitting and absorbing disks, $I_0$ is the
brightness in the center of the O star disk, $i$ is the orbital inclination,
$x$ is the limb darkening coefficient of the O star. The functions
$K_1(\xi,\Delta,R_{\rm O})$, $K_2(\xi,\Delta,R_{\rm O})$
describe the shape of the overlapping disk area within the eclipses (their
analytical form is given in \cite{cher75}, \cite{gon85}). $I_c(\xi)$ and
$I_a(\xi)$ are unknown functions which describe brightness and absorption
distributions, respectively, over the WR disk. The latter can be represented
in the form $I_a(\xi)=1-e^{-\tau(\xi)}$ where $\tau(\xi)$ is the optical
depth along the line of sight in the extended WR atmosphere at the impact
distance $\xi$.

Thus, the primary and secondary minima provide two integral equations to be
solved, their left-hand sides containing observational light losses and the
right-hand ones involving the integrals of the unknown functions described
above. As a matter of fact, the number of parameters to be determined is not
as large as it may seem. The brightness distribution over the disk of a
main-sequence star can be satisfactorily described by a linear
limb-darkening law. The problem thus involves only two respective
parameters, the linear limb-darkening coefficient, $x$, and the brightness
in the center of the O star disk, $I_0$. Linear darkening coefficients for
early main-sequence stars are well known from stellar atmosphere models;
$x\,=\,0.3$ for an O5 V star in the optical range. The central brightness
$I_0$ is determined from the third, algebraic, equation which connects
the relative luminosities of the components. Thus, the problem involves two
unknown functions, $I_a(\xi)$ and $I_c(\xi)$, together with two unknown
geometrical parameters, the orbital inclination angle $i$, and the radius of the O star,
$R_{\rm O}$. It is important to note that one can only restore the
brightness distribution over the whole WR disk including its central parts,
if the center of the WR disk is covered by the O star disk edge during the
secondary eclipse. This condition can be expressed as an equation for the
method applicability, $\cos(i)\,\le\,R_{\rm O}$. We cannot know by default
if this condition is fulfilled. There is, however, a simple and sufficient
condition for the method applicability \cite{cher73}: if the light loss in
the lowest point of the secondary minimum (O star is in front) is equal to,
or more than a half of the relative observational luminosity of the WR
component, then clearly at least a half of the WR disk is invisible. If so,
the center of the WR disk is certainly covered by the center of the O star.

The integral equations (1a,b) are Fredholm equations of the 1st kind which
are known to be ill-posed problems \cite{gon85}. Such problems could not be
solved without some a-priori information concerning the behavior of the
unknown function. The simplest example of such information is a parametric
representation of the unknown function. For example, if a binary system
consists of two main-sequence stars, the function $I_c(\xi)$ can be
represented by a linear limb-darkening law and $I_a(\xi)$ as being equal to
$1$ when $\xi\,<\,R_{\rm WR}$ and $0$ when $\xi\,\ge\,R_{\rm WR}$. Then the
problem could be reduced to a set of classical equations of the theory of
eclipsing binary systems.

For a WR star we can not use a linear limb-darkening law for
$I_c(\xi)$. Since the wind of the WR star is semitransparent, a simple
appproximation for $I_a(\xi)$ given above is of no use either. It was shown
in \cite{tikhonov43} that an inverse problem can be considered correct if
its solution is searched on a compact set of functions. To solve such an
inverse problem, standard methods can be used. Thus, to choose a type of a
priori information acceptable in our case, one could ask a question: what
minimal assumptions about the unknown functions can be done to make the
problem correct in the classical sense? In \cite{tikhonov43}, \cite{gon85}
it was shown that these assumptions can be very common in nature. In out
case, the minimal requirements to the unknown $I_c(\xi)$ and
$I_a(\xi)$ are these: the functions must be restricted, non-negative and
monotonically non-increasing.

In practice these requirements tend to lead to solutions of low physical sense
(for example \cite{gon85}, \cite{ant01}) when distribution functions found
are of a stepped form. Hence, in our work
\cite{ant01} dealing with light curve analysis of the eclipsing binary
WN5\,+\,O6\,V V444\,Cyg more strict requirements on the functions
$I_c(\xi)$ and $I_a(\xi)$ were used. These functions were supposed to belong to
a set of convexo-concave, non-increasing,  monotonic and non-negative
functions. A convex part of the functions corresponds to the core of a WR
star which contains most of stellar mass while a concave one describes the
extended photosphere and the atmosphere of the WR star. A position of an
inflection point is a free parameter of the problem. Such a priori
information allows peculiarities of the WR star and its extended atmosphere
to be taken into account in the least model-dependent way. The same
requirements on the unknown functions are used in the current work. One
additional condition is added, however. The radius of the opaque part of the
WR core (where $I_a(\xi)=1.0$) is required to be equal to, or larger than
the halfwidth of the function $I_c(\xi)$ at the half of its maximum. In
doing so, the widths of the functions $I_c(\xi)$ and
$I_a(\xi)$ are made consistent. Indeed, the central part of the WR disk
should be absolutely opaque. If there were no such a restriction, a
better description of the primary minimum could be achieved for a function
$I_a(\xi)$ hot having any flat part in the region of small $\xi$. Typically,
this happens at large $i$. A ``wide'' flat-topped function $I_a(\xi)$
deepens the model primary minimum, and the only way for the minimization
algorithm to reduce the divergence with the observational light curve is to
decrease the width of the flat top in $I_a(\xi)$. Because of that, the
algorithm chooses as the ``best'' solution the one which does not have a
flat top at all, i.e. the WR disk becomes semitransparent right from its
center. The new restriction allows one to overcome this problem.

Numerical solution of the problem (1) for a given pair of geometrical parameters
$i$, $R_{\rm O}$ is done through discretization of the problem. The unknown
continuous functions $I_c(\xi)$ and $I_a(\xi)$ are represented as discrete
functions defined in the knots of a rather dense grid over $\xi$, the
integrals in (1) are replaced by sums, and the solution of the integral
equations in (1) is found from residual minimization, the residual being a
weighted sum of difference squares between the left and the right hands in
equations (1a,b). First the equation for the secondary minimum is solved,
then $I_0$ is determined from the normalization equation, and after that the
equation is solved which describes the primary minimum.

Since the problem involves only two independent geometrical parameters (the
orbital inclination and the O star radius) it is possible to look for the
best solution by solving the problem on a reasonable grid on these
parameters. This method has the advantage of avoiding trapping in a local
minimum. A normalized residual is defined as a weighted sum of squares of
deviations between a model light curve and the observational one divided by
the sum of weights. Since the problem is solved for two minima the resulting
residual is a sum of two residuals obtained for the corresponding minima.

The main disadvantage of the method is that it is impossible to estimate how
reliable is the resulting solution using standard statistical criteria of
$\chi^2$ type. The matter is, the parameters of the problem are related in
such a way that a real number of degrees of freedom could not be evaluated
reliably. In theoretical works \cite{tikhonov43}, \cite{gon85} there is
nothing but a mention that, while solving problem (1) on a {\em compact}
class of functions, an approximate solution converges to the true one when
observational errors decrease. In our case, we are dealing with a discrete
light curve and a discrete representation for the unknown functions. The
latters, generally speaking, not necessarily have to be smooth. Accuracy of
the observational light curve as well as the number of data points it
consists of are the two factors which define the quality of the approximate
solution. Indeed, if there was only one observational point in each eclipse
the functions $I_c(\xi)$ and $I_a(\xi)$ could vary within huge limits still
perfectly satisfying equations (1) and our a priori restrictions. For
example, they can be mere constants.
 
\begin{figure*}
\epsfig{file=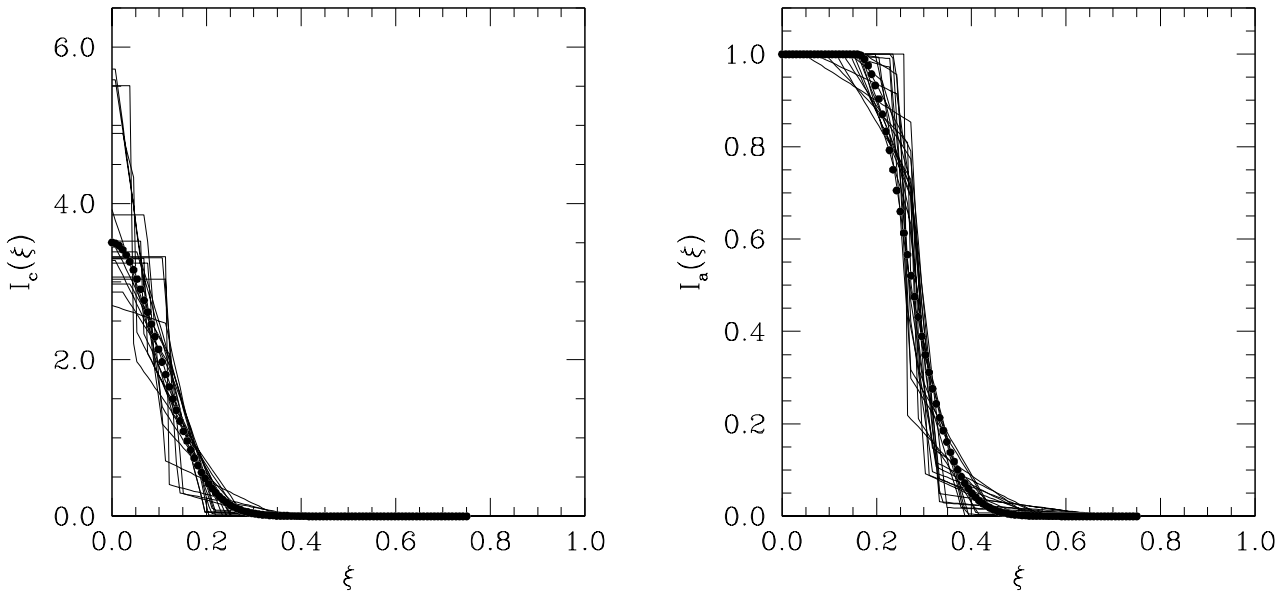,width=\textwidth}

\caption{Results of a test solution. An input light curve for given
orbital parameters and functions $I_c(\xi)$, $I_a(\xi)$ was calculated on
the same orbital phases as the real observed light curve of BAT99-129. Random
deviations were then added to this ``ideal'' light curve distributed
according to the normal law with root-mean-square deviation 0.003. In this
way 100 simulated light curves containg random errors have been obtained and
solved. Black points stand for the true functions $I_c(\xi)$ and $I_a(\xi)$.
Solid lines represent the solutions for these functions for different
realizations of the perturbed input light curve (20 solutions are shown to
avoid encumbering the visualization).}

\label{fig-sim}
\end{figure*}

In our work \cite{ant07} we carried out many numerical experiments solving
artificially generated light curves by means of the above described method
in order to find out the effect of the mentioned factors on the solutions
obtained. Here we briefly summarize the conclusions of this work in relation
to an artificially created light curve that resembles the observed light
curve of BAT99-129 (Fig.\,\ref{fig-sim}). The geometrical parameters of the
system are determined very reliably, the deviations from actual values do
not exceed 2-5 per cent. The radius of the opaque WR star core is also
determined quite reliably (from the shape of the functions $I_c(\xi)$ and
$I_a(\xi)$). The value of $I_c(0)$ that is necessary to determine the
brightness temperature of the WR star (see below) may have very large error.
It is more likely to overestimate a real temperature than to underestimate it. A
typical spread of the model $I_c(0)$ is $-40\%\div +100\%$ realtive to the
``true'' temperature.

Using the function $I_c(\xi)$ obtained in solving (1) and assigning the
O star a value of effective temperature according to its spectral type, it
is possible to determine the brightness temperature of the WR star in the
central parts of its disk that is characteristic of the core temperature
\cite{gon78}:

\begin{equation}
\label{eq2}
\begin{array}{rl}
T_b(\xi,\lambda)=&\frac{1.44}{\displaystyle\lambda\ln(\displaystyle\frac{1}{A}e^{1.44/\lambda{T}}+1-\displaystyle\frac{1}{A})}, \\[+10mm]
A=&\displaystyle\frac{\pi{R_{\rm O}^2\left[1-\frac{x(\lambda)}{3}\right]I_c(\xi,\lambda)}}{\displaystyle1-L_{\rm WR}(\lambda)}
\end{array}
\end{equation}

Note that the so obtained value of $T_b(\xi,\lambda)$ does not depend on
interstellar absorption since the O star is used as a comparison star and
the whole method of determination is differential. (For details see
\cite{cher75}, \cite{cher84}, \cite{gon85}).

\section{Solution of the Light Curve of BAT99-129}

The light loss in the center of  the secondary minimum is $1-l_2(0)=0.14$.
The relative observational luminosity of the WR component is
$L^{obs}_{\rm WR}=0.25$. Thus, the sufficient condition for applicability of
the method is fulfilled: $1-l_2(0)>\frac{1}{2}L^{obs}_{\rm WR}$.

The light of BAT99-129 between the eclipses exhibit regular variability with
the amplitude around 1\,--\,2\,\% (Fig.\,\ref{fig-macho_lc}). This means
that the binary system components are not quite spherical. This may also
testify about mutual heating of the components. Our technique cannot take
these factors into consideration. In the classical theory of eclipsing
binary systems a rectified light curve is solved in such cases
\cite{russell52}. The standard rectification formulae (for example, for the
model of rotating ellipsoids) cannot be applied to BAT99-129 (as one of the
components is a WR star). We can derive, however, an empirical rectification
formula approximating the light of the binary system outside of eclipses:
$l=a_0+a_1\cdot{\cos\theta}+a_2\cdot\cos{2\theta}$, where
$\theta$ is a phase angle in radians \cite{russell52}. A special question is
revealing the phases which would signify the beginning and termination of
the eclipses. Following recommendations from \cite{russell52}, we have
choosen the phase intervals 0.0\,--\,0.1 for the primary minimum and
0.4\,--\,0.5 for the secondary one. Approximation of the light between
eclipses by the least-square method yields $a_0=0.99311$, $a_1=-0.01108$,
$a_2=-0.00462$.

To better judge the reliability of our results we applied the above
described technique to both non-rectified and rectified mean light curves of
BAT99-129. The solutions are presented below.
 
\subsection{Non-rectified Light Curve}

\begin{figure}
\epsfig{file=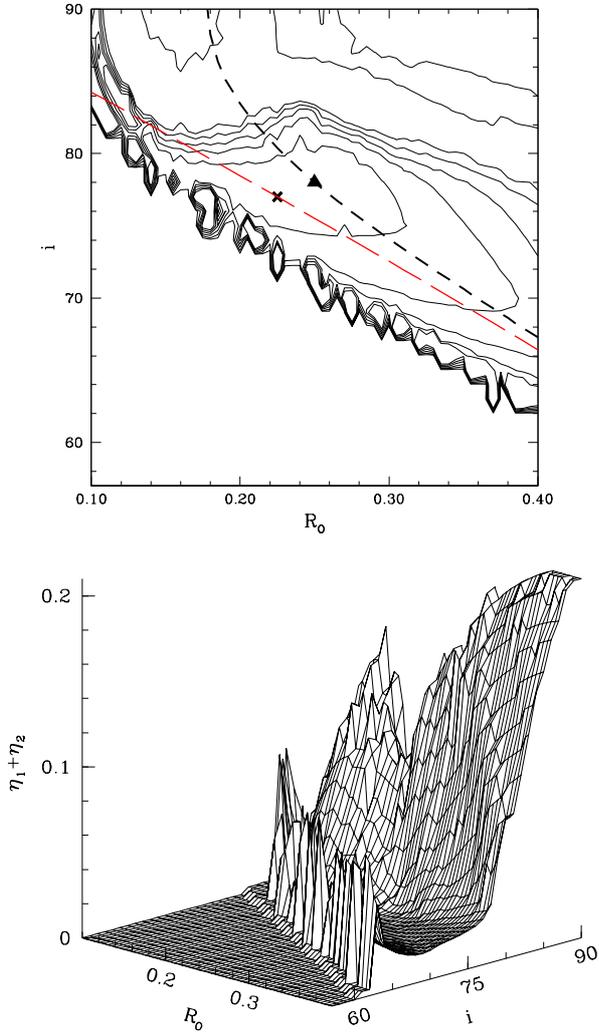,width=\linewidth}

\caption{The surface of the normalized residuals (bottom). The isolevels
of this surface are shown on the top. The long-dashed line shows the
boundary corresponding to the equation $cos(i)\,=\,R_{\rm O}$. The
short-dashed line corresponds to a fixed relative WR luminosity
$L_{\rm WR}=0.25$. The cross marks the absolute minimum of the residuals.
The triange marks the minimum along the line of the fixed WR luminosity. The
isolevels which are the closest to the optimal solution have the values of
$0.010$, $0.015$.}

\label{fig-norect_eta}
\end{figure}

\begin{figure}
\epsfig{file=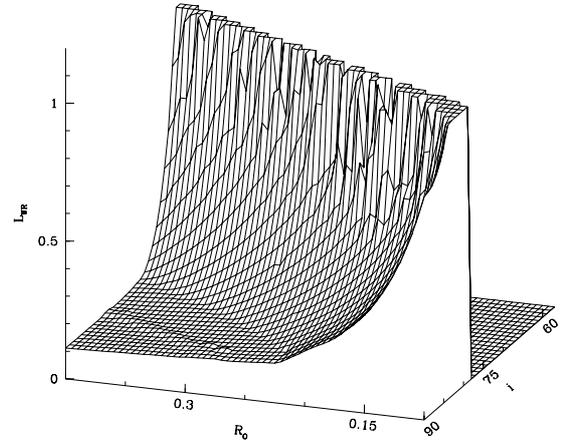,width=\linewidth}

\caption{The relative luminosity of the WR component versus geometrical parameters;
the non-rectified light curve.}

\label{fig-norect_lwr}
\end{figure}

The mean random error of a normal point in an averaged non-rectified light
curve is $0.0030$. The surface of residuals for the non-rectified light curve
is shown in Fig.\,\ref{fig-norect_eta}. Each pair of geometrical parameters
$R_{\rm O}$, $i$ corresponds to a specific relative luminosity of the
components (see Fig.\,\ref{fig-norect_lwr}). The absolute minimum of the
residuals $(\eta_1+\eta_2)_{min}=0.0079$ is achieved at
$R_{\rm O}=0.225$, $i=77^\circ$ (marked by the cross in
Fig.\ref{fig-norect_eta}. In this case, the relative luminosity of the WR
component is $L_{\rm WR}=0.377$.

\begin{figure}
\epsfig{file=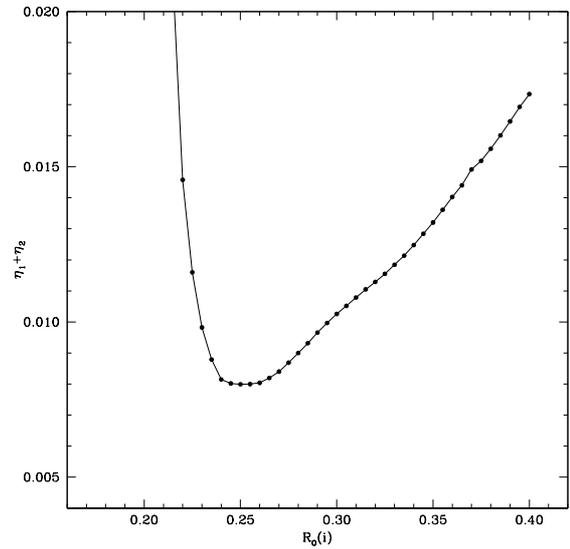,width=\linewidth}

\caption{The residuals along the line of fixed luminosity $L_{\rm
WR}\,=\,0.25$ for the non-rectified light curve.}

\label{fig-norect_etaroi}
\end{figure}

As was mentioned in Section 2, the observational ratio of the binary
component luminosities is $L^{obs}_{\rm WR}/L^{obs}_{\rm
O}\,=\,0.34\pm{0.20}$. This value corresponds to the relative WR component
luminosity of WR $L^{obs}_{\rm WR}=0.25\pm{0.11}$. The geometrical
parameters corresponding to this luminosity ratio are shown in Fig.
\ref{fig-norect_eta} as a solid broken line with short dashes. The minimum
of the residual $\eta_1+\eta_2$ along this line (marked by the triangle) is
equal to $0.0080$ (see also Fig.\,\ref{fig-norect_etaroi}). The minimum
corresponds to the parameters $R_{\rm O}=0.25$, $i=78^\circ.013$. The
residual in this point is practically identical to the residual in the
absolute minimum and only slightly exceeds the doubled mean error in the
observational light curve. Partly, this may be due to the fact that our
model is not able to describe small regular variability of BAT99-129 outside
the eclipses. Hence, we have chosen, as a final solution, the parameters for
the fixed WR component luminosity, $L_{\rm WR}=0.25$: $i=78^\circ$, $R_{\rm
O}=0.25$. Since the minimal residual exceeds the error in the observational
light curve, it is not possible to formally estimate (even roghly)
uncertainties of these parameters. The absolute orbital separation
corresponding to this orbital inclination is $28.5\,R_\odot$.

\begin{figure}
\epsfig{file=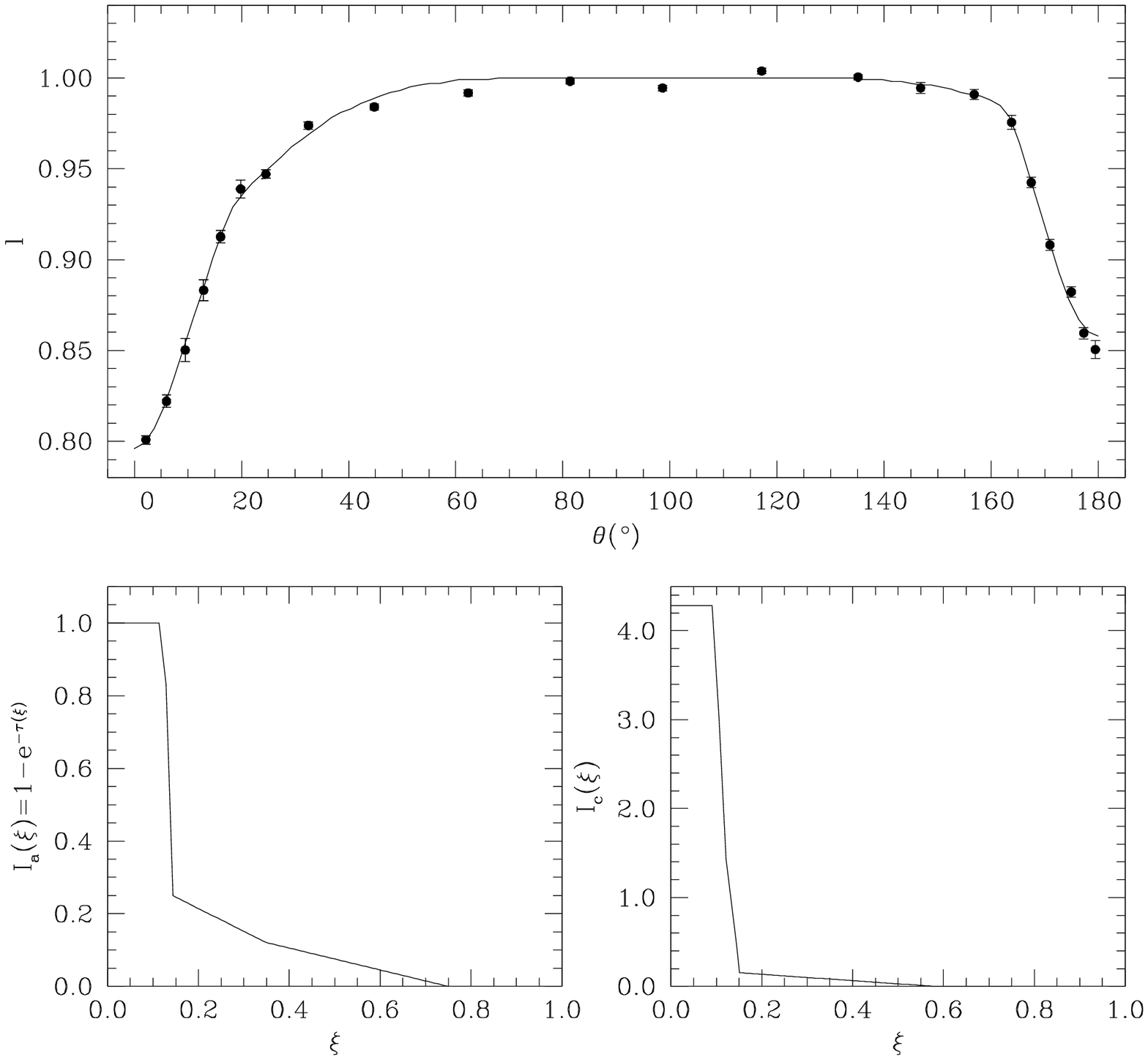,width=\linewidth}

\caption{The solution of the non-rectified light curve for the optimal
values of geometrical parameters.}

\label{fig-norect_sol}
\end{figure}

A model light curve and the functions $I_c(\xi)$, $I_a(\xi)$ for the chosen
solution are shown in Fig.\,\ref{fig-norect_sol}. The behavior of the
function $I_a(\xi)$ indicates that the radius of the opaque WR core is 0.13
of the orbital size ($3.7\,R_\odot$).

Using the value of $I_c(0)=4.28$, the parameters of the O star described in
Section 2 as well as those obtained in the result of solving the problem
(1), and the relative luminosity of the WR star, one can determine the
brightness temperature in the center of the WR star disk by means of formula
(2). This temperature specifies the core WR temperature. The wavelength in
(2) is chosen to be equal to the mean wavelength of the MACHO spectral
range, 6150\,\AA. The brightness temperature of the WR star core proves out
to be $\sim{43\,000}$\,K.

\subsection{Rectified Light Curve}

\begin{figure}
\epsfig{file=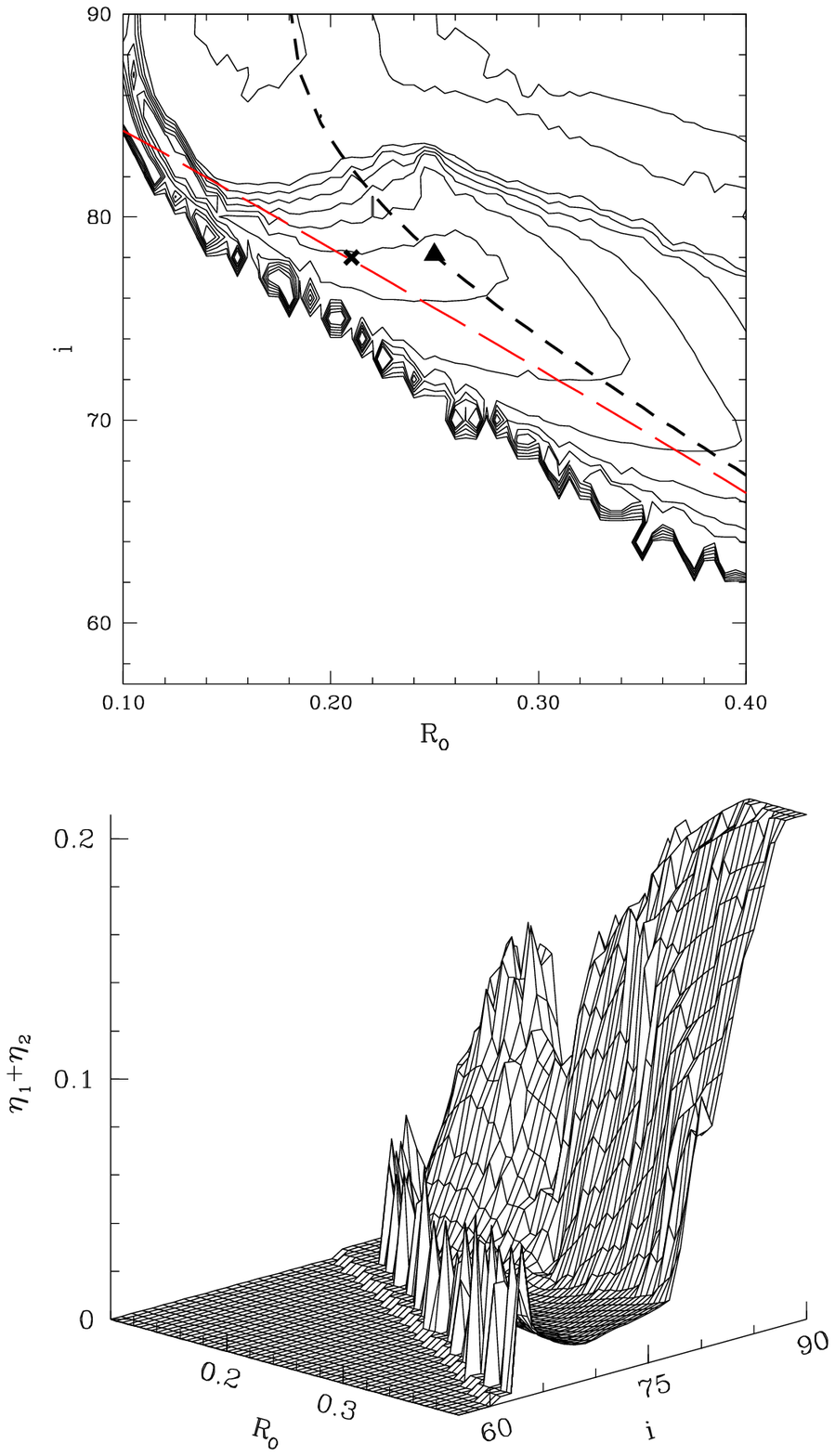,width=\linewidth}

\caption{The surface of residuals for the rectified light curve. The
designations are the same as in Fig.\,\ref{fig-norect_eta}. The isolevels
which are the closest to the optimal solutions have the values of 0.006 and
0.010.}

\label{fig-rect_eta}
\end{figure}

\begin{figure}
\epsfig{file=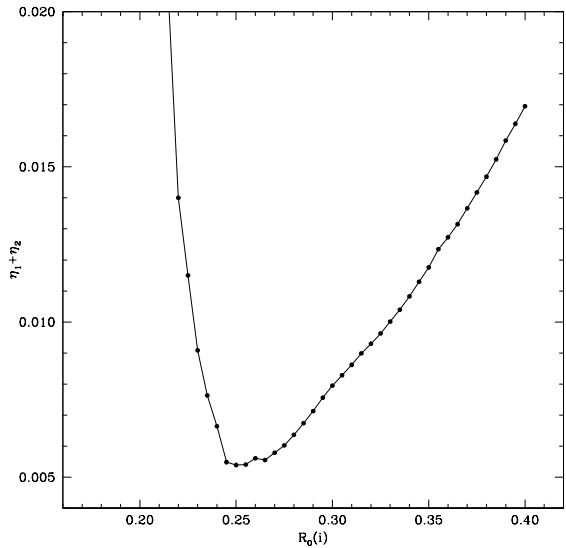,width=\linewidth}

\caption{The residuals along the line of the fixed luminosity $L_{\rm
WR}\,=\,0.25$ for the rectified light curve.}

\label{fig-rect_etaroi}
\end{figure}

The surface of residuals for the rectified light curve is shown in
Fig.\,\ref{fig-rect_eta}. The notations are the same as in
Fig.\,\ref{fig-norect_eta}. As in the case of the non-rectified light curve,
the geometrical parameters of the solutions chosen according to the absolute
residual minimum and according to the minimum which corresponds to the
fixed WR component luminosity, are slightly different. In both cases,
the residuals are practically identical, the absolute minimum being
$0.0053$, and the residual at the fixed WR star luminosity, $0.0054$ (see
also Fig.\,\ref{fig-rect_etaroi}). These values are somewhat lower than the
doubled mean error of a data point in the mean light curve. As a final
solution, we have chosen the one which corresponds to the fixed WR
luminosity, $L_{\rm WR}=0.25$: $i=78^\circ.14$, $R_{\rm O}=0.25$.


In Section 2 we emphasised the difficulties to estimate a probable
uncertainty of the solution. Very roughly, uncertainty in the
geometrical parameters could be estimated by the isolevel which
corresponds to the observational error. In doing so, the observational error
of $L^{obs}_{\rm WR}$ has also to be taken into account. The intervals
$76^\circ.0\leq{i}\leq{79^\circ.5}$ and $0.21\leq{R_{\rm O}}\leq{0.30}$ in
the $R_{\rm O}$, $i$ plane define the area where the normalized residual
does not exceed the observational error and $L_{\rm WR}$ is constrained
within the limits corresponding to the observational luminosity and its
error.

\begin{figure}
\epsfig{file=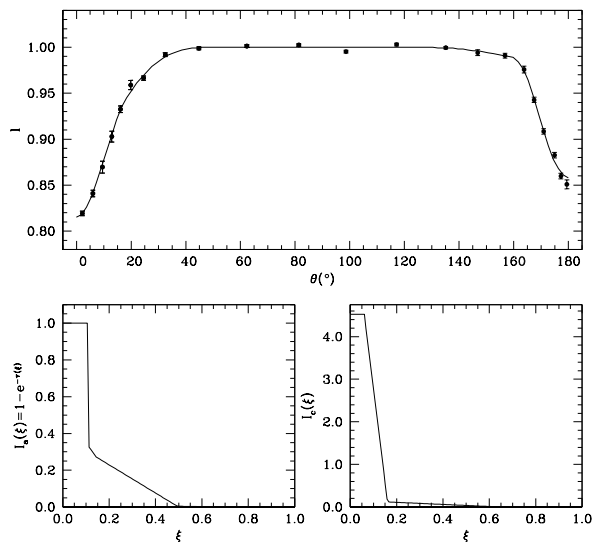,width=\linewidth}

\caption{The rectified light curve solution for the optimal geometrical
parameters.}

\label{fig-rect_sol}
\end{figure}

The model light curve and the obtained functions $I_c(\xi)$, $I_a(\xi)$ for
the chosen optimal solution are shown in Fig.\,\ref{fig-rect_sol}. The
opaque core radius of the WR star is 0.12 of the orbital size
($3.4\,R_\odot$). In this case, the WR brightness temperature is equal to
$45\,000$\,K ($I_c(0)=4.52$).

One can estimate an uncertainty in the opaque core radius of the WR star and
in the brightness temperature in the center of its disc. Let's find the
$i - R_{\rm O}$ dependency for two fixed values of the WR star relative
luminosity: 0.36 and 0.14, which are the extreme values of the $L_{\rm WR}$
uncertainty interval. Let's find the residual minimum along each $i - R_{\rm
O}$ curve similarly to how this was done for $L_{\rm WR}=0.25$. It turnes
out that the two so obtained points on the $R_{\rm O}$, $i$ plane are
located at the opposite sides of the error box given in the previous
subsection. The solutions in these points yield the following values of the
WR core radius and the WR temperature: 0.1 and 0.15 ($2.8$ and
$4.3\,R_\odot$), $49\,000$\,K and $43\,000$\,K, for $L_{\rm WR}=0.14$ and
$0.36$ respectively. These limits can be accepted as rought estimates of
the probable confidence intervals in the corresponding parameters.

\section{System Parameters and its Evolutionary Status}

Based on our analysis of the MACHO BAT99-129 light curve the following
conclusions concerning the parameters of the system and its components can
be derived:

\begin{enumerate}

\item The most plausible orbital inclination is $78^\circ$, a probable
uncertainty interval being $76^\circ\,-\,79^\circ.5$. Using the results from
\cite{foellmi06}: $a\sin\,i=27.9\,R_\odot$, $M_{\rm
WR}\sin^3i=14\pm{2}M_\odot$, $M_{\rm O}\sin^3i=23\pm{2}M_\odot$, we can
estimate the absolute parameters of the system: $a=28.5\,R_\odot$, $M_{\rm
WR}=15\pm{2}\,M_\odot$, $M_{\rm O}=24.6\pm{2}\,M_\odot$.

\item The most plausible O star radius is 0.25 of the distance between the
components, that is $7.12\,R_\odot$. A probable uncertainty interval for
this radius is 0.21\,-\,0.30, that is $6.0\,-\,8.6\,R_\odot$ in absolute
units. At an effective temperature of the O star $43\,000$\,K its bolometric
luminosity is $\lg{L_{\rm O}/L_\odot}\,=\,5.18$.

\item The most plausible opaque core radius of the WR star is 0.12 of the
orbital separation ($3.4\,R_\odot$). A probable error is $0.10\,-\,0.15$
($2.8\,-\,4.3\,R_\odot$).

\item The brightness temperature of the WR core is $45\,000$\,K (its
uncertainty is $43\,000\,-\,49\,000$\,K). Let us recall that this
uncertainty is a rather formal one. Simulations show that a probable
uncertainty can be up to some dozen per cent. It is more likely to
overestimate the temperature than to underestimate it. The optimal
temperature is in agreement with the results obtained from spectral modeling
of isolated WR stars \cite{hamann04} and with our results obtained for
the V444\,Cyg \cite{ant01} system.

\end{enumerate}

The function $I_a(\xi)$ can potentially be used to restore the spatial structure
of the extended WR atmosphere \cite{ant01}. This problem is also an
ill--posed one. Such analysis was not carried out in present work because of
great uncertainty of its results. To perform such analysis, high-accuracy
narrow-band observations of BAT99-129 in optical and IR continuum are
desirable.

A small radius of the opaque WR star core ($3.4\,R_\odot$), its mass equal
to $15\,M_\odot$, as well as high brightness temperature of the core are
indicative of an evolved massive hot star. The initially more massive
component of a binary system can transform into a WR star in two ways (see,
for example, \cite{mas88}). It can lose the most part of its mass via
stellar wind. The material leaves the system, the accretion onto the
companion being negligible. Both components evolve as if they were mere
single stars. In the second scenario a more massive star fills its critical
Roche lobe as it evolves and the so called initial mass exchange occurs in
the system. The material flows from the more massive component to its
companion. In wide binary systems mass exchange is conservative and stellar
material does not escape the system. In very close systems, a less massive
component can also fill its Roche lobe after it accumulates enough material.
Such a system becomes contact for a while. A common envelope forms around
the system and some material can leave it taking away part of the system's
angular momentum.

The authors of \cite{foellmi06} argue that the evolution of BAT99-129 went
through a contact stage. First, the WR wind mass loss rate is too small to
explain the whole mass loss from the system on evolutionary timescale.
Second, the period of the system (hence, its orbital separation) is very
small. If mass exchange occured without the contact phase (a conservative
case), the period and the orbital separation of the system would increase.
The mass exchange scenario without contact phase implies that the initial
period of the system should have been shorter than 2 days which is almost
improbable. In contact mass exchange, however, the orbital period and size
are decreasing. Since the authors of \cite{foellmi06} did not have precise
information on the orbital inclination of the system they suggested
``typical'' masses for the binary system components according to the
spectral type -- mass dependence.

We have reliably estimated the orbital inclination of BAT99-129 which allows
us to get some additional insight into the evolutionary scenario for the
system. Let us suppose that initial conservative mass exchange took place in
the system without a contact phase. In this case, the mass of a helium
remnant (WR star) evolved from a massive star is related to the initial mass
of its progenitor via the approximate relation \cite{tutukov73}:
$M_{\rm He}\simeq{0.1\cdot{M^{1.4}_{\rm O}}}$. Using this formula leads to
the initial mass of the WR star progenitor being equal to $36\,M_\odot$.
Thus, the mass loss by the WR star during its evolution is about
$21\,M_\odot$. Since mass exchange is supposed to be conservative the
initial mass of the O component should have been only $4\,M_\odot$ which is
highly unlikely. These considerations show that the mass exchange could not
be conservative, that is, some part of stellar material should have escaped
the system. WR star formation according to the first scenario (mass loss due
to stellar wind) appears to be unlikely, too. To transform into a WR star,
the progenitor's mass should have been more than $25\,M_\odot$ taking into
account its rotation and metallicity $Z=0.008$ \cite{meynet05}. However, if
a star had such initial mass it inevitably would pass through a supergiant
stage before passing to the WR stage. In a close binary system such as
BAT99-129, this would lead to contact mass exchange. A WR progenitor with
greater mass (for example, $\geq{50}\,M_\odot$) which would imply the
absence of a supergiant stage in the course of evolution is less probable.
In that case the system would inevitably lose a considerable part of
material to form a circumstellar nebula. Nebula lines in the spectrum of
BAT99-129, however, have not been found \cite{foellmi06}. It looks liely
that the initial mass of the progenitor amounted to no more than
$\sim\,40\,M_\odot$. A part of this mass has escaped the system at the stage
of contact exchange. Nevertheless, this portion of mass is relatively small
and has not been observed. To answer the question decisively one should
obtain accurate spectra of the system with high spectral resolution and
signal-to-noise ratio.

\begin{figure}
\epsfig{file=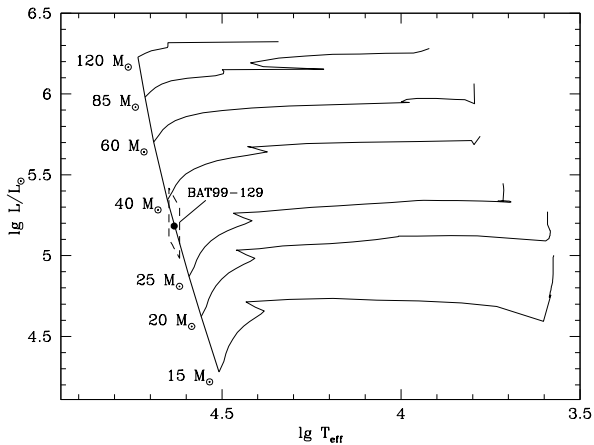,width=\linewidth}

\caption{ The evolutionary diagram for massive stars \cite{meynet94}.
Metallicity is $Z=0.008$. To avoid cluttering, evolutionary tracks for the
most massive stars are not shown completely. The bold point stands for the O
component in the binary BAT99-129. The dashed line shows a probable error
zone (see the text).}

\label{fig-evolution}
\end{figure}

While there appears to be some confidence in the parameters and the
evolutionary status of the WR component, the situation with the O component
is far less certain. In Fig.\,\ref{fig-evolution} the position of the O star
is shown on an evolutionary diagram for massive stars \cite{meynet94}. A
dotted area stands for a probable error zone (see below). Evolutionary
tracks corresponding to the metallicity $Z=0.008$ of the Large Magellanic
Cloud (LMC) are used. Stellar rotation is not taken into consideration. The
models with rotation are considered in other works (see, for example,
\cite{meynet00}). Unfortunately, evolutionary tracks with rotation have not
been published for the LMC metallicity. In any case, a position of the Zero
Age Main Sequence (ZAMS) on the evolutionary diagram does not depend
strongly on rotation \cite{meynet00}. The star is located exactly on ZAMS.
Its temperature and luminosity correspond to a ZAMS star with the mass of
about $35\,M_\odot$, while the observational mass is only $25\,M_\odot$.
Recall that the observational mass has been obtained using a rather reliable
spectral estimate $M_{\rm O}\sin^3i=23\,M_\odot$ \cite{foellmi06} and our
estimate for the orbital inclination. If the O star mass were $35\,M_\odot$,
the orbital inclination would have to be about 60 degrees which is absolutely
incompatible with the results of our analysis.

\begin{figure}
\epsfig{file=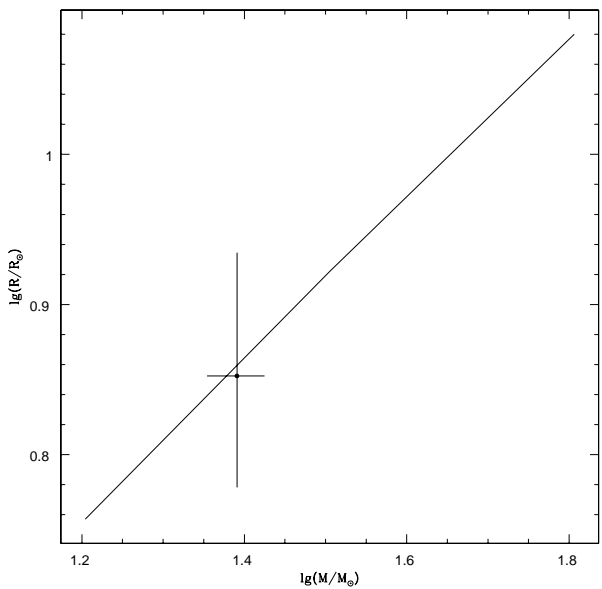,width=\linewidth}

\caption{The theoretical dependence of mass versus radius for ZAMS stars.
The position of the O component in the binary BAT99-129 is shown.}

\label{fig-mr}
\end{figure}

According to \cite{meynet94}, a ZAMS star with $25\,M_\odot$ and $Z=0.008$
has the temperature $39\,000$\,K and the luminosity $lg{L_{\rm
O}/L_\odot}=4.87$. Obviously, there exists an excess in the O star
luminosity and its temperature. The mass and radius of the O component
correspond perfectly to a theoretical mass--radius dependence for ZAMS stars
(\cite{tutukov73}). The position of the O component in BAT99-129 on this
dependence is shown in Fig.\,\ref{fig-mr}. The extra luminosity and
temperature of the O component can be explained in three possible ways:

\begin{enumerate}

\item Large errors in our $T_{eff}$ and $L_{\rm O}$. The error zone given in
Fig.\,\ref{fig-evolution} has been obtained in the following way. The error
of $T_{eff}$ is set to $1500\,$K (this error defines the positions of the
left and right borders of the error zone). This value is a typical {\em intrinsic}
accuracy of temperature calibration which is set by comparison of atmosphere
models with spectra of real stars. Unfortunately, the models used by
different authors yield systematic discrepancies which may considerably
exceed the intrinsic accuracy of the individual models. As noted in
\cite{massey05}, this may be partly due to the fact that different models
emphasis different spectral lines and spectral regions. In other words,
contemporary models of stellar atmospheres are not quite adequate with real
physics. As a result, ``the scale of absolute effective O star temperatures
is likely to be revised in future'' \cite{massey05}. Moreover, temperatures
of different stars which belong to the same spectral subtype may differ
significantly (even if a single atmospheric model is used
\cite{herrero03}). Hence, the error zone shown in Fig.\,\ref{fig-evolution}
should be considered as minimal. On the other hand, the shortcomings of modern
atmospheric models are likely to affect the ZAMS position on the
evolutionary diagram as well. Thus, a {\em relative} position of the O star
on this diagram would possibly not change too much when using another
temperature calibration.

The borders of the error zone defined by the error in the O star radius (the
upper and the bottom borders) correspond to $6.0$ and
$8.6\,R_\odot$. Although it has been stressed above that we cannot provide
truly sound statistical error estimates of the model parameters, it looks
very unlikely that the radius of the O star is outside the above limits.

Concluding, one can state that, while large errors in the parameters found
cannot be formally ruled out as a reason for the ``strange'' position of the
O star on the evolutionary diagram, this reason looks not very likely.

\item The O component may belong rather to the O6 than O5 subtype.
This possibility is mentioned in \cite{foellmi06}. The temperature and
luminosity of an O6 component appear to be considerably lower than those of
O5. To coincide the position of the O component in  BAT99-129 system in
Fig.\ref{fig-evolution} with the initial point on the track for a  
$25\,M_\odot$ star, its temperature should be $T_{\rm O,eff}=39\,000$\,K,
and the radius $R_{\rm O}=6.0\,R_\odot$. The temperature $39\,000$\,K is
typical for the Galactic O6 stars and seems a bit low for the LMC ones
(however, see our notes above). The value of the radius, while being right
on the bottom edge of a probable error box, seems to be somewhat too low.
Finally, this opportunity can be verified when high-accuracy spectra of
BAT99-129 are available which would allow one to refine the spectral
classification of the O star.
 
\item In the process of the initial mass exchange, outer layers of the O
component are enriched with the nuclear burning products formed in the WR
progenitor. This enforces mixing of stellar material \cite{vanbev94}. The
process is further supported by increased rotation velocity of the O
component due to transfer of the angular momentum from the primary
component. Spectral line widths of the O component in BAT99-129 reveal its
possible asynchronous rotation \cite{foellmi06}. Thus, this component may be
a chemically homogeneous star enriched by the CNO-cycle products. The
mass-radius dependence shown in Fig.\,\ref{fig-mr} is valid for any
chemically homogeneous star \cite{tutukov73}, which may explain the position
of the O component in this Figure. On the other hand, the anomalous chemical
composition may be in charge of this component luminosity excess. Indeed, as
was shown in \cite{mas88}, in massive MS stars in which the radiation
pressure is dominant, the radius weakly depends on the chemical composition.
At the same time, other conditions being equal, the luminosity is
proportional to the ratio of the mean molecular weight to
$1+X$ where $X$ is the relative hydrogen abundance. Since the molecular
weight increases and the value of $X$ decreases in the result of the initial
mass exchange, the luminosity should somewhat rise compared to a star of
the same mass and of the solar chemical composition.

\end{enumerate}

\section{Conclusion} 

Wide-band observations of the eclipsing binary system BAT99-129 in optical
continuum are interpreted in terms of an inverse problem solution. Specific
a priori information was used to constrain the unknown functions $I_c(\xi)$
and $I_a(\xi)$ which represent brightness and absorption distributions over
the WR disk: they are convex-concave, monotonously non-increasing and
non-negative functions. A convex part corresponds to the WR star core while
a concave part describes the extended photosphere and atmosphere of the WR
star. The method used to solve the inverse ill-posed problem and to restore
the unknown functions does not allow one to formally estimate the parameter
errors. To estimate the order of actual errors, and also the influence of
the rectification procedure on the results, both the original non-rectified
and the rectified light curves were solved. As a result of such analysis the
orbital inclination of the binary system was estimated and the parameters of
the components were determined along with their likely uncertainties.

The analysis allows one to make some conclusions concerning the evolutionary
status of the binary. The progenitors of the binary components likely had
masses around 20--40$\,M_\odot$ and the system appears to have passed
through a contact phase when it lost part of stellar material in the process
of the initial mass exchange. The parameters of the WR star in the binary
are typical for such objects. The O star shows some excess in its luminosity
and temperature. The excess can be due to various reasons, most likely ones
are uncertainty in spectral type determination and/or changes in the stellar
chemical composition during the initial mass exchange.

There are only a few eclipsing binary systems of the WR+O kind where the
parameters of the WR stars can be reliably determined. BAT99-129 is one of
such systems. Its further observations in the optical and X-ray domains are
highly desirable.

The authors wish to thank Dr. A.V.Tutukov, the referee of the work, for his
valuable remarks. The data from the MACHO project base have been used. The
work has been supported by the Russian Foundation for Basic Research( grant
N 05-02-17489).

\end{document}